\documentclass[10pt,twosides,a4paper]{book}
\usepackage{amssymb}
\usepackage[pdftex]{graphicx}
\usepackage{latexsym}
\usepackage[small]{caption2}
\usepackage{amsmath}
\usepackage[english]{babel}
\usepackage{babel}
\usepackage{amsmath,amsthm}
\usepackage{rotating}
\usepackage{multirow}
\usepackage{graphicx}
\usepackage{tocloft}
\usepackage{bbm}
\usepackage{dsfont}
\usepackage{amsfonts}
\usepackage{accents}
\usepackage{pdfpages}
\usepackage[thinc]{esdiff}
\usepackage{hyperref}
\usepackage[numbers,sort&compress,square]{natbib}
\usepackage{natbib}
\usepackage{pdfsync}
\usepackage{fancyhdr}
\usepackage{scalerel,stackengine}
\usepackage{fancyhdr}
\renewcommand{\cftchappresnum}{CHAPTER }
\newlength{\mylen}
 \fancypagestyle{plain}{}

\begin{document}


\chapter*{Graviton induced decoherence of a composite particle}

\markboth{Graviton induced decoherence of a composite particle}{T. H. Moreira and L. C. Céleri}

\noindent{\large \textbf{T. H. Moreira$^{1,a}$ and L. C. Céleri$^{1,b}$}}

\vspace{3mm}

\noindent{$^1$ QPequi Group, Institute of Physics, Federal University of Goi\'as, Goi\^ania, Goi\'as, Brazil}


\noindent{\texttt{$^a$thiagohenriquemoreira@discente.ufg.br; $^{b}$lucas@qpequi.com}}


\section{Introduction}

It is safe to say that one of the greatest open problems in physics is the lack of a full, experimentally confirmed, quantum description of gravity, for which even its necessity is still under debate. Nevertheless, whatever complete description we may have for quantum gravity for all energy scales (if needed), one may expect to recover, in the weak field limit, the usual quantum field theoretic description. To be more precise, the weak field limit involves a metric expansion around some classical solution to Einstein's field equation, where only the dynamics of the small perturbation are considered. Classically, this leads to the description of gravitational waves. Since this is mathematically equivalent to describing a field over a fixed background spacetime structure, the quantization procedure follows straightforwardly in complete analogy with the quantization of the other known interactions in the Standard Model of particle physics. The quantum of the metric perturbation field is a particle we call \emph{graviton}.

Although the possibility of single graviton detection is under debate~\cite{Dyson_2013}, there have been some proposals for probing the quantum nature of the gravitational field that do not involve directly detecting the graviton, like gravity-induced entanglement for instance~\cite{Bose2017,Marletto2017,Danielson_2022,Christodoulou2023,Christodoulou2023b}. In particular, another approach was taken by Parikh, Wilczek, and Zahariade (PWZ)~\cite{Parikh_2021,Parikh2021}, and later generalized by Cho and Hu~\cite{Cho2022} (see also Ref.~\cite{Kanno2021}), in which the basic idea is to detect the quantum fluctuations of spacetime as stochastic noise entering in a Langevin-like equation for the classical geodesic deviation of two test masses, in the same spirit as the quantum Brownian motion.

Among other topics of interest in the intersection between gravitational physics and quantum mechanics, some attention has been given to gravitational decoherence, by which we mean loss of quantum coherence of a system that is in some way related to gravitational interactions. Gravitational decoherence is a broad term that includes, for instance, the decoherence of a quantum nonrelativistic particle in a superposition of positions due to the coupling with a bath of gravitons~\cite{Kanno2021}, decoherence of quantum matter modeled by a scalar field coupled with a weak quantum gravitational field~\cite{Blencowe_2013,Anastopoulos_2013}, and also the loss of quantum coherence in the center-of-mass wave function of a composite particle induced by the coupling with its internal degrees of freedom through gravitational time dilation~\cite{Pikovski2015,Pikovski2017}, among many other approaches along this direction~\cite{Bassi_2017}.

In these scenarios, we investigate the loss of quantum coherence in a superposition of the external degrees of freedom (DoFs) of a quantum system which is induced by the coupling with its own internal DoFs and with spacetime quantum fluctuations, namely a bath of gravitons. Following the PWZ approach, we use the Feynman-Vernon influence functional technique~\cite{Feynman1963,Feynman2010,Calzetta2008} to obtain the reduced density matrix describing the non-unitary evolution of the external variables of the system. Due to the universal aspect of gravity, the field also couples with the internal degrees of freedom, meaning that the two environments (internal DoFs and gravitons) also interact with each other and the solution to the problem cannot be obtained by simply adding the individual influences of each of them.

Here we use the metric convention $\eta_{\mu\nu}=\textrm{diag}(-1,+1,+1,+1)$. Additionally, unless explicitly stated otherwise, we work in units such that $\hbar=c=k_B=G=1$, where $\hbar$ is the reduced Planck constant, $c$ is the speed of light in vacuum, $k_B$ is the Boltzmann constant, and $G$ is the gravitational constant.

\section{Quantum matter in a bath of gravitons}

\subsection{The classical action}

It follows from the equivalence principle that a single particle is not enough to probe the effects of a gravitational field~\cite{Carroll}. Let us then start by writing the classical action of a weak gravitational field coupled to a pair of free-falling massive particles. The total action takes the form
\begin{equation} \label{Total-action-matter+h}
    S=S_{\rm matter}+S_{\rm EH},
\end{equation}
where the first term describes the pair of free-falling test masses while the second one is the Einstein-Hilbert action which describes the dynamics of the metric field $g_{\mu\nu}$.

By denoting the spacetime coordinates of each particle by $\zeta^\mu$ and $\xi^\mu$, the matter action can be written as
\begin{equation} \label{Matter-action}
    S_{\rm matter}=-M_0\int \textrm{d} t\,\sqrt{-g_{\mu\nu}\dot{\zeta}^\mu\dot{\zeta}^\nu}+\int \textrm{d} t\,L_{\rm rest}\sqrt{-g_{\mu\nu}\dot{\xi}^\mu\dot{\xi}^\nu},
\end{equation}
where $M_0$ is the mass of the first particle (with coordinates $\zeta^\mu$) and $L_{\rm rest}$ denotes the rest Lagrangian of the second particle (with coordinates $\xi^\mu$). In writing down the action~\eqref{Matter-action} we are assuming that the first particle has no (dynamical) internal DoFs. For the second particle, the rest Lagrangian takes the form
\begin{equation} \label{L_rest}
    L_{\rm rest}(\varrho,\dot{\varrho}\,\bar{t})=-m_0+\lambda\mathcal{L}(\varrho,\dot{\varrho}\,\bar{t}).
\end{equation}
In this equation, $m_0$ is the mass of the particle, while $\mathcal{L}(\varrho,\dot{\varrho}\,\bar{t})$ describes its internal degrees of freedom with coordinate $\varrho$ and generalized velocity $\dot{\varrho}=\textrm{d}\varrho/\textrm{d}t$. We have also defined $\bar{t}=\textrm{d}t/\textrm{d}\tau$, with $\tau$ being the particle's proper time, and $\lambda$ is a dimensionless control parameter. Additionally, we consider that the first particle is on-shell with worldline $\zeta_0^\mu(t)$ and $M_0 \gg L_{\rm rest}$. Furthermore, we take this particle at rest at the origin of our coordinate system, $\zeta_{0}^\mu(t)=t\,\delta_0^\mu$, such that the coordinate time $t$ is interpreted as its proper time. Under these assumptions, the first term in the action~\eqref{Matter-action} essentially has no dynamics, and we will refer to the composite particle simply as the system from now on~\cite{Parikh2021,Moreira2023}.

In the context we described above, it becomes appropriate to think of $(t,\xi^i)$ as the Fermi normal coordinates defined with respect to the worldline of the heavier particle~\cite{Manasse1963}. In these coordinates, we can write the metric components as
\begin{equation*}
    \begin{split}
        g_{00}(t,\xi^i)&=-1-R_{i0j0}(t,0)\xi^i\xi^j+O(\xi^3), \\
        g_{0i}(t,\xi^i)&=-\frac{2}{3}R_{0jik}(t,0)\xi^j\xi^k+O(\xi^3), \\
        g_{ij}(t,\xi^i)&=\delta_{ij}-\frac{1}{3}R_{ikjl}(t,0)\xi^k\xi^l+O(\xi^3),
    \end{split}
\end{equation*}
where $R_{\mu\nu\rho\sigma}$ is the Riemann curvature tensor. We neglected higher-order terms since we are describing a nonrelativistic, slowly moving, composite particle.

In our parameterization $\xi^0(t)=t$, thus resulting
\begin{align} \label{dot-tau}
    \sqrt{-g_{\mu\nu}(\xi)\dot{\xi}^\mu\dot{\xi}^\nu}&\simeq\sqrt{1+R_{i0j0}(t,0)\xi^i\xi^j-\delta_{ij}\dot{\xi}^i\dot{\xi}^j} \nonumber \\
    &\simeq1-\frac{1}{2}\delta_{ij}\dot{\xi}^i\dot{\xi}^j+\frac{1}{2}R_{i0j0}(t,0)\xi^i\xi^j,
\end{align}
in the nonrelativistic limit.

Now, since we want to consider the interaction with weak gravitational radiation, we expand the metric field as
\begin{equation} \label{g=n+h}
    g_{\mu\nu}=\eta_{\mu\nu}+h_{\mu\nu},
\end{equation}
where $\eta_{\mu\nu}$ is the Minkowski metric as usual and we assume $\left| h_{\mu\nu}\right| \ll1$. We also take advantage of the gauge invariance of linearized gravity~\cite{Carroll,Weinberg2013} and choose to work in the transverse-traceless (TT) gauge, defined by the relations  
\begin{equation} \label{TT-gauge}
\begin{split}
    \Bar{h}^{0\mu}=0, \\
    \eta^{\mu\nu}\Bar{h}_{\mu\nu}=0, \\
    \partial^i \Bar{h}_{ij}=0.
\end{split}
\end{equation}
The Riemann tensor in the TT gauge of linearized gravity takes the form~\cite{Parikh_2021,Moreira2023}
\begin{equation} \label{Riemann-tensor-TT-gauge-2}
    R_{i0j0}(t,0)=-\frac{1}{2}\ddot{\Bar{h}}_{ij}(t,0).
\end{equation}
Plugging Eqs.~\eqref{L_rest},~\eqref{dot-tau} and~\eqref{Riemann-tensor-TT-gauge-2} into the matter action~\eqref{Matter-action} results
\begin{equation} \label{Matter-action-2}
\begin{split}
    S_{\rm matter}&=\frac{1}{2}\int \textrm{d} t\,m_0\delta_{ij}\dot{\xi}^i\dot{\xi}^j+\lambda\int \textrm{d} t\,\mathcal{L}(\varrho,\dot{\varrho}\Bar{t})-\frac{1}{2}\lambda\int \textrm{d} t\,\mathcal{L}(\varrho,\dot{\varrho}\Bar{t})\delta_{ij}\dot{\xi}^i\dot{\xi}^j \\
    &\hspace{0,5cm}+\frac{1}{4}\int \textrm{d} t\,\left[ m_0-\lambda\mathcal{L}(\varrho,\dot{\varrho}\Bar{t})\right] \ddot{\Bar{h}}_{ij}(t,0)\xi^i\xi^j,
\end{split}
\end{equation}
where we dropped non-dynamical terms. Note that the first three terms in Eq.~\eqref{Matter-action-2} describe the system alone, which involves a free term for the external DoFs, another one for the internal DoFs and the interaction between them. The last term describes the interaction between the system and the weak gravitational field.

Now, for the metric field, the Einstein-Hilbert action in the TT gauge can be written as
\begin{equation}\label{Eistein-Hilbert-action-TT-gauge}
    S_{\rm EH}=-\frac{1}{64\pi}\int \textrm{d}^4x\,\partial_\mu\Bar{h}_{ij}\partial^\mu\Bar{h}^{ij},
\end{equation}
from which we immediately obtain the wave equation $\Box\Bar{h}_{ij}=0$. We may then express the gravitational waves in terms of its two physical degrees of freedom, corresponding to the two polarization components $s=+,\times$, by means of a Fourier transform,
\begin{equation} \label{Fourier-transform}
    \Bar{h}_{ij}(t,\mathbf{x})=\int \textrm{d}^3k\,\sum_s\epsilon_{ij}^{s}(\mathbf{k})q_s(t,\mathbf{k})e^{i\mathbf{k}\cdot\mathbf{x}},
\end{equation}
with $\epsilon_{ij}^{s}$ denoting the polarization tensor which satisfies the normalization $\epsilon_{ij}^s(\mathbf{k})\epsilon^{ij}_{s'}(\mathbf{k})=2\delta^s_{s'}$, the transversality $k^i\epsilon_{ij}^{s}(\mathbf{k})=0$ and the traceless $\delta^{ij}\epsilon_{ij}^{s}(\mathbf{k})=0$ conditions. Additionally, the reality of $\Bar{h}_{ij}$ implies that
\begin{equation} \label{Implication-of-reality-of-h-ij}
    \epsilon_{ij}^{s}(\mathbf{k})q_s^*(t,\mathbf{k})=\epsilon_{ij}^{s}(-\mathbf{k})q_s(t,-\mathbf{k}).
\end{equation}

By plugging Eq.~\eqref{Fourier-transform} into Eq.~\eqref{Eistein-Hilbert-action-TT-gauge} we obtain
\begin{align*}
    S_{\rm EH}&=-\frac{1}{64\pi}\int \textrm{d}^4x\,\int \textrm{d}^3k\,\textrm{d}^3k'\,\sum_{s,s'}\epsilon_{ij}^s(\mathbf{k})\epsilon^{ij}_{s'}(\mathbf{k}') \nonumber \\
    &\hspace{1cm}\times\left[ -\dot{q}_s(t,\mathbf{k})\dot{q}_{s'}(t,-\mathbf{k}')-\mathbf{k}\cdot\mathbf{k}'q_s(t,\mathbf{k})q_{s'}(t,\mathbf{k}')\right] e^{i(\mathbf{k}+\mathbf{k}')\cdot\mathbf{x}} \nonumber \\
    &=-\frac{1}{64\pi}\int \textrm{d}t\int \textrm{d}^3k\,\textrm{d}^3k'\,\sum_{s,s'}\epsilon_{ij}^s(\mathbf{k})\epsilon^{ij}_{s'}(-\mathbf{k}') \nonumber \\
    &\hspace{1cm}\times\left[ -\dot{q}_s(t,\mathbf{k})\dot{q}_{s'}^*(t,-\mathbf{k}')-\mathbf{k}\cdot\mathbf{k}'q_s(t,\mathbf{k})q_{s'}^*(t,-\mathbf{k}')\right] \int \textrm{d}^3x\,e^{i(\mathbf{k}+\mathbf{k}')\cdot\mathbf{x}} \nonumber \\
    &=\frac{\pi^2}{8}\int \textrm{d}t\int \textrm{d}^3k\,\sum_s\left[ \left| \dot{q}_s(t,\mathbf{k})\right| ^2-\mathbf{k}^2\left| q_s(t,\mathbf{k})\right| ^2\right] ,
\end{align*}
where we used Eq.~\eqref{Implication-of-reality-of-h-ij} as well as the identity
\begin{equation*}
    \delta^3(\mathbf{k}+\mathbf{k}')=\frac{1}{(2\pi)^3}\int \textrm{d}^3x\,e^{i(\mathbf{k}+\mathbf{k}')\cdot\mathbf{x}}.
\end{equation*}

It is easy to see that only the real part of the wave amplitude couples to the matter degrees of freedom, allowing us to take those to be real. Explicitly,
\begin{align*}
    \ddot{\Bar{h}}_{ij}(t,0)\xi^i\xi^j&=\int \textrm{d}^3k\,\sum_s\epsilon_{ij}^s(\mathbf{k})\ddot{q}_s(t,\mathbf{k})\xi^i\xi^j \\
    &=\frac{1}{2}\int \textrm{d}^3k\,\sum_s\epsilon_{ij}^s(\mathbf{k})[\ddot{q}_s(t,\mathbf{k})+{\ddot{q}_s}^*(t,\mathbf{k})]\xi^i\xi^j \\
    &=\int \textrm{d}^3k\,\sum_s\epsilon_{ij}^s(\mathbf{k})\textrm{Re}[\ddot{q}_s(t,\mathbf{k})]\xi^i\xi^j,
\end{align*}
which follows from Eqs.~\eqref{Fourier-transform} and~\eqref{Implication-of-reality-of-h-ij}. The total action~\eqref{Total-action-matter+h} can then be written as
\begin{align} \label{Total-action-matter+h-2}
    S&=\frac{1}{2}\int \textrm{d} t\,m_0\delta_{ij}\dot{\xi}^i\dot{\xi}^j+\lambda\int \textrm{d} t\,\mathcal{L}(\varrho,\dot{\varrho}\Bar{t})-\frac{1}{2}\lambda\int \textrm{d} t\,\mathcal{L}(\varrho,\dot{\varrho}\Bar{t})\delta_{ij}\dot{\xi}^i\dot{\xi}^j \nonumber \\
    &\hspace{0.5cm}+\frac{\pi^{2}}{8}\int \textrm{d} t\int \textrm{d}^3k\,\sum_s\left[ \dot{q}^2_s(t,\mathbf{k})-\mathbf{k}^2q_s^2(t,\mathbf{k})\right] \nonumber \\
    &\hspace{0.5cm}+\frac{m_0}{4}\int \textrm{d} t\,\int \textrm{d}^3k\,\sum_sq_s(t,\mathbf{k})\,X^s(t,\mathbf{k}),
\end{align}
where we have defined
\begin{equation} \label{X(t)-definition}
    X^s(t,\mathbf{k})\equiv\frac{\textrm{d}^2}{\textrm{d}t^2}\left\{ \epsilon_{ij}^s(\mathbf{k})\xi^i\xi^j\left[ 1-\frac{\lambda}{m_0}\mathcal{L}(\varrho,\dot{\varrho}\bar{t})\right] \right\} .
\end{equation}

The action~\eqref{Total-action-matter+h-2} describes the interaction of a system, whose dynamics involves both external and internal DoFs, with the classical gravitational radiation. We next proceed to give a complete quantum mechanical description of the total system. The quantization of the metric perturbation field can be done by promoting the field amplitudes to operators in Hilbert space in the usual canonical approach to Quantum Field Theory. However, since our ultimate goal is to describe the system variables alone, the weak quantized gravitational field shall be treated as an environment, for which case the Feynman-Vernon influence functional approach to open quantum systems becomes much more appealing.

\subsection{Gravitational influence functional} \label{Sec:Gravitational-influence-functional}

Assuming all measurements to be made in the system alone, we are not interested in the graviton's final state. Therefore, we must integrate over the gravitational variables by means of a partial trace, such that the system becomes effectively open and the weak quantum gravitational field is viewed as an environment. This description is accomplished by using the Feynman-Vernon influence functional~\cite{Feynman1963,Feynman2010}, which is written as~\cite{Cho2022}
\begin{equation*}
    \mathcal{F}[x,x']=e^{iS_{\rm IF}[x,x',t]},
\end{equation*}
where $S_{\rm IF}$ is called the influence action and $x$ and $x'$ denote two different histories of the system variables (the reader is referred to Ref.~\cite{Calzetta2008} for a review). In terms of the influence functional, the time evolution of the system reduced density matrix $\rho_r$, which is obtained from the total density matrix by taking the partial trace with respect to the environment variables, is given by
\begin{equation*}
    \rho_r(x,x',t)=\int \textrm{d} x_0\textrm{d} x_0'\,\mathcal{J}_r(x,x',t|x_0,x_0',0)\rho_{\rm sys}(x_0,x_0',0),
\end{equation*}
where $\rho_r(x,x',t)=\langle x|\rho_r(t)|x'\rangle$ and the evolution operator for the reduced density matrix reads
\begin{equation*}
    \mathcal{J}_r(x,x',t|x_0,x_0',0)\equiv\int\displaylimits_{\substack{x(0)\,=\,x_0 \\ x'(0)\,=\,x_0'}}^{\substack{x(t)\,=\,x \\ x'(t)\,=\,x'}}\mathcal{D}x\mathcal{D}x'\,e^{i(S_{\rm sys}[x]-S_{\rm sys}[x'])}\mathcal{F}[x,x'],
\end{equation*}
with $S_{\rm sys}$ denoting the part of the action describing the system alone and $\mathcal{D}x$ denoting the path integral measure. The exponentiated system action is responsible for the unitary process, while the influence functional describes a non-unitary evolution in general.

The influence action encodes all influence of the environment on the system. Since each graviton mode (and polarization) is independent of all the others, they can be treated separately in such a way that the total influence action is the sum of the actions corresponding to each mode (and polarization). This is possible because the total influence functional for a system coupled with statistically and dynamically independent environments is simply the product of each individual influence functional. Furthermore, we note that the free environment action (the Einstein-Hilbert action) is quadratic in the field amplitudes, and the coupling with the system variable $X^s(t)$, Eq.~\eqref{X(t)-definition}, is linear. This is then a special case of the linear coupling model, for which the influence action is exactly found to be of the form~\cite{Calzetta2008,Cho2022,Moreira2023}
\begin{equation} \label{S_IF}
\begin{split}
    S_{\rm IF}[x,x']=\int \textrm{d} t\textrm{d} t'\left\{ \frac{1}{2}[x_{ij}(t)-x'_{ij}(t)]D^{ijkl}_{\rm g}(t,t')[x_{kl}(t')+x'_{kl}(t')] \right. \\
    \left. +\frac{i}{2}[x_{ij}(t)-x'_{ij}(t)]N^{ijkl}_{\rm g}(t,t')[x_{kl}(t')-x'_{kl}(t')]\right\} ,
\end{split}
\end{equation}
where
\begin{equation} \label{x-ij(t)-definition}
    x_{ij}(t)\equiv\xi_i(t)\xi_j(t)\left[ 1-\frac{\lambda}{m_0}\mathcal{L}(\varrho,\dot{\varrho}\bar{t})\right] ,
\end{equation}
and
\begin{subequations}
    \begin{equation}
        D^{ijkl}_{\rm g}(t,t')=i\left( \frac{m_0}{4}\right) ^2\frac{\textrm{d}^2}{\textrm{d}t^2}\frac{\textrm{d}^2}{\textrm{d}{t'}^2}\sum_s\int \textrm{d}^3k\,\epsilon^{ij}_s(\mathbf{k})\epsilon^{kl}_s(\mathbf{k})\langle\left[ q_s(t,\mathbf{k}),q_s(t',\mathbf{k})\right] \rangle_{\rm g}\theta(t-t'),
    \end{equation}
    \begin{equation} \label{Noise-kernel-definition}
        N^{ijkl}_{\rm g}(t,t')=\frac{1}{2}\left( \frac{m_0}{4}\right) ^2\frac{\textrm{d}^2}{\textrm{d}t^2}\frac{\textrm{d}^2}{\textrm{d}{t'}^2}\sum_s\int \textrm{d}^3k\,\epsilon^{ij}_s(\mathbf{k})\epsilon^{kl}_s(\mathbf{k})\langle\left\{ q_s(t,\mathbf{k}),q_s(t',\mathbf{k})\right\} \rangle_{\rm g},
    \end{equation}
\end{subequations}
are the (gravitational) dissipation and noise kernels respectively~\cite{Cho2022,Calzetta2008}, with the $q$'s standing for position operators in the Heisenberg picture. Here $[\cdot,\cdot]$ and $\{\cdot,\cdot\}$ denote the commutator and anti-commutator of operators, respectively, and $\theta(x)$ is the Heaviside step function. The expectation values with the subscript 'g' are computed with respect to the initial state of the gravitons. The term proportional to the noise kernel can be expressed in terms of a stochastic variable $\mathcal{N}_{ij}(t)$ using the Gaussian functional identity~\cite{Cho2022}
\begin{equation*}
\begin{split}
    &e^{-\frac{1}{2}\int \textrm{d} t\textrm{d} t'\,[x_{ij}(t)-x'_{ij}(t)]N^{ijkl}_{\rm g}(t,t')[x_{kl}(t')-x'_{kl}(t')]} \\
    &\hspace{1cm}=\mathcal{C}\int\mathcal{D}\mathcal{N}\,e^{-\frac{1}{2}\int \textrm{d} t\textrm{d} t'\,\mathcal{N}_{ij}(t)(N_g^{-1})^{ijkl}(t,t')\mathcal{N}_{kl}(t')}e^{-i\int \textrm{d} t\,\mathcal{N}^{ij}(t)[x_{ij}(t)-x'_{ij}(t)]},
\end{split}
\end{equation*}
where $\mathcal{C}$ is a normalization constant and $\mathcal{D}\mathcal{N}$ denotes the path integral measure for the stochastic variable $\mathcal{N}_{ij}(t)$. Stochastic averages are computed using the Gaussian probability density
\begin{equation*}
    P[\mathcal{N}]=\mathcal{C}\,e^{-\frac{1}{2}\int \textrm{d} t\textrm{d} t'\,\mathcal{N}_{ij}(t)(N_g^{-1})^{ijkl}(t,t')\mathcal{N}_{kl}(t')}.
\end{equation*}
For example, the average and the two-point correlation function are
\begin{subequations}  \label{Stochastic-averages}
\begin{equation}
    \langle\mathcal{N}^{ij}(t)\rangle_{\rm sto}=\int\mathcal{D}\mathcal{N}\,P[\mathcal{N}]\mathcal{N}^{ij}(t)=0,
\end{equation}
\begin{equation}
    \langle\mathcal{N}^{ij}(t)\mathcal{N}^{kl}(t')\rangle_{\rm sto}=\int\mathcal{D}\mathcal{N}\,P[\mathcal{N}]\mathcal{N}^{ij}(t)\mathcal{N}^{kl}(t')=N_{\rm g}^{ijkl}(t,t').
\end{equation}
\end{subequations}

The gravitational influence functional then becomes
\begin{equation} \label{Influence-functional-final}
\begin{split}
    e^{iS_{\rm IF}[x,x']}&=\int\mathcal{D}\mathcal{N}\,P[\mathcal{N}]\,e^{-i\int \textrm{d} t\,\mathcal{N}^{ij}(t)[x_{ij}(t)-x'_{ij}(t)]} \\
    &\hspace{0.5cm}\times e^{\frac{i}{2}\int \textrm{d} t\textrm{d} t'[x_{ij}(t)-x'_{ij}(t)]D^{ijkl}_{\rm g}(t,t')[x_{kl}(t')+x'_{kl}(t')]}.
\end{split}
\end{equation}
Let us now note that the term involving the dissipation kernel in Eq.~\eqref{Influence-functional-final} is of order $O(\xi^4)$ since $x_{ij}$ is already of order $O(\xi^2)$. Thus the leading order contribution comes from the term involving the noise kernel and we may approximate Eq.~\eqref{Influence-functional-final} as
\begin{equation} \label{Influence-functional-final-approximated}
    e^{iS_{\rm IF}[x,x']}\simeq\int\mathcal{D}\mathcal{N}\,P[\mathcal{N}]\,e^{-i\int \textrm{d} t\,\mathcal{N}^{ij}(t)[x_{ij}(t)-x'_{ij}(t)]}.
\end{equation}
This means that we are only considering the influence of the gravitational field encoded into the noise kernel, for which one can derive an explicit expression by considering a specific initial graviton state.

\subsection{Gravitational noise kernel}

In what follows, we shall consider the position operators to be independent of the polarizations and of the direction of $\mathbf{k}$, that is, $q_s(t,\mathbf{k})=q_\omega(t)$, $\omega=|\mathbf{k}|$. This means that the angular integral we need to compute in order to obtain the noise kernel is
\begin{equation*}
    \int \textrm{d} \Omega\sum_s\epsilon_s^{ij}(\mathbf{k})\epsilon_s^{kl}(\mathbf{k}).
\end{equation*}
The polarization tensors, for the 'plus' and 'cross' polarizations~\cite{Carroll}, can be written as
\begin{equation*}
\begin{split}
    \epsilon_{ij}^{+}&=\hat{\epsilon}_i^{(1)}\hat{\epsilon}_j^{(1)}-\hat{\epsilon}_i^{(2)}\hat{\epsilon}_j^{(2)} \\
    \epsilon_{ij}^{\times}&=\hat{\epsilon}_i^{(1)}\hat{\epsilon}_j^{(2)}+\hat{\epsilon}_i^{(2)}\hat{\epsilon}_j^{(1)},
\end{split}
\end{equation*}
where the spatial polarization unit vectors $\hat{\epsilon}^{(1)}$ and $\hat{\epsilon}^{(2)}$ are orthogonal to the direction of propagation $\hat{k}=\mathbf{k}/|\mathbf{k}|$. They satisfy
\begin{equation*}
    \hat{\epsilon}_i^{(1)}\hat{\epsilon}_j^{(1)}+\hat{\epsilon}_i^{(2)}\hat{\epsilon}_j^{(2)}=\delta_{ij}-\hat{k}_i\hat{k}_j\equiv P_{ij}.
\end{equation*}
Then, a straightforward calculation yields
\begin{equation*}
    \sum_s\epsilon_s^{ij}\epsilon_s^{kl}=P^{ik}P^{jl}+P^{il}P^{jk}-P^{ij}P^{kl}.
\end{equation*}

We can use the following integration over solid angles in three spatial dimensions:
\begin{equation*}
    \int \textrm{d}\Omega\,\hat{k}_i\hat{k}_j=\frac{4\pi}{3}\delta_{ij},
\end{equation*}
\begin{equation*}
    \int \textrm{d}\Omega\,\hat{k}_i\hat{k}_j\hat{k}_k\hat{k}_l=\frac{4\pi}{15}(\delta_{ij}\delta_{kl}+\delta_{ik}\delta_{jl}+\delta_{il}\delta_{jk}),
\end{equation*}
to obtain
\begin{equation*}
    \int \textrm{d}\Omega\,P_{ij}P_{kl}=\frac{8\pi}{5}\delta_{ij}\delta_{kl}+\frac{4\pi}{15}(\delta_{ij}\delta_{jl}+\delta_{il}\delta_{jk}).
\end{equation*}
We finally find
\begin{subequations} \label{Angular-integral}
\begin{equation}
    \int \textrm{d}\Omega\sum_s\epsilon_s^{ij}(\mathbf{k})\epsilon_s^{kl}(\mathbf{k})=\frac{8\pi}{15}\mathcal{P}^{ijkl},
\end{equation}
where
\begin{equation} \label{Pijkl}
    \mathcal{P}^{ijkl}\equiv3(\delta^{ik}\delta^{jl}+\delta^{il}\delta^{jk})-2\delta^{ij}\delta^{kl}.
\end{equation}
\end{subequations}

In order to compute the noise kernel~\eqref{Noise-kernel-definition}, we will need to evaluate the expectation value of the anti-commutator of position operators at different times with respect to the graviton's initial state. Since the free Lagrangian for each mode of the gravitational field is the one of a harmonic oscillator, we may write the position operators in the Heisenberg picture as
\begin{equation*}
    q_\omega(t)=\sqrt{\frac{1}{2m\omega}}(a_\omega e^{-i\omega t}+a_\omega^\dagger e^{i\omega t}),
\end{equation*}
where the $a$'s ($a^\dagger$'s) are annihilation (creation) operators satisfying the usual commutation relations
\begin{equation*}
\begin{split}
    \left[ a_\omega,a_{\omega'}\right] &=\left[ a_\omega^\dagger,a_{\omega'}^\dagger\right] =0, \\
    \left[ a_\omega,a_{\omega'}^\dagger\right] &=\delta_{\omega\omega'}.
\end{split}
\end{equation*}
A direct calculation then yields
\begin{equation} \label{Expval-anticomm}
\begin{split}
    \langle\left\{ q_\omega(t),q_\omega(t')\right\} \rangle _{\rm g}&=\frac{2}{m\omega^2}\langle H_\omega\rangle_{\rm g}\cos\omega(t-t') \\
    &\hspace{0.2cm}+\frac{1}{m\omega}\left[ \langle a_\omega^2\rangle_{\rm g}e^{-i\omega(t+t')}+\langle(a^\dagger_\omega)^2\rangle_{\rm g}e^{i\omega(t+t')}\right] ,
\end{split}
\end{equation}
where
\begin{equation*}
    H_\omega=\omega\left( a_\omega^\dagger a_\omega+\frac{1}{2}\right) =\frac{\omega}{2}\left\{ a_\omega,a_\omega^\dagger\right\} 
\end{equation*}
is the free Hamiltonian operator of the harmonic oscillator with frequency $\omega$.

If the initial state is the vacuum state, we have $\langle a_\omega^2\rangle_{\rm g}=\langle(a^\dagger_\omega)^2\rangle_{\rm g}=0$ and $\langle H_\omega\rangle _{\rm g}=\omega/2$, so that
\begin{equation*}
    \langle\left\{ q_\omega(t),q_\omega(t')\right\} \rangle_{\rm g}=\frac{1}{m\omega}\cos\omega(t-t').
\end{equation*}
Then, from Eqs.~\eqref{Noise-kernel-definition} and~\eqref{Angular-integral} we have
\begin{equation*}
\begin{split}
    N_{\textrm{g (vac)}}^{ijkl}(t,t')&=\frac{m_0^2}{15\pi}\mathcal{P}^{ijkl}\int_0^\infty \textrm{d}\omega\,\omega^5\cos\omega(t-t').
\end{split}
\end{equation*}
This integral over $\omega$ diverges in its upper limit, requiring us to introduce the cutoff $\Lambda_{\rm g}$, such that~\cite{Cho2022,Kanno2021}
\begin{equation*}
    \int_0^{\Lambda_{\rm g}}d\omega\,\omega^5\cos\omega(t-t')=\Lambda_{\rm g}^6\,F[\Lambda_{\rm g}(t-t')],
\end{equation*}
where
\begin{align} \label{F(x)}
    F(x)&\equiv\frac{1}{x^6}\int_0^x\textrm{d} y\,y^5\cos y \nonumber \\
    &=\frac{1}{x^6}\left[ (5x^4-60x^2+120)\cos x+x(x^4-20x^2+120)\sin x-120\right] .
\end{align}

At last, the noise kernel for the Minkowski vacuum initial state reads
\begin{equation} \label{Noise-kernel-Minkowski-vacuum}
    N_{\textrm{g (vac)}}^{ijkl}(t,t')=\frac{m_0^2\Lambda_{\rm g}^6}{15\pi}\mathcal{P}^{ijkl}F[\Lambda_{\rm g}(t-t')].
\end{equation}

\subsection{The external DoFs density matrix}

Having explicitly computed the gravitons' influence through the noise kernel for specific initial states, let us now turn to the evolution of the system density matrix. The particle total density matrix (external plus internal degrees of freedom) at time $t$ is given by
\begin{equation} \label{Total-density-matrix-of-system}
\begin{split}
    \rho_{\rm sys}(\xi\varrho,\xi'\varrho,t)=\int \textrm{d}\xi(0)\textrm{d}\xi'(0)\textrm{d}\varrho(0)\textrm{d}\varrho'(0)\,\rho_{\rm sys}(\xi(0)\varrho(0),\xi'(0)\varrho'(0),0) \\
    \times\mathcal{J}(\xi\varrho,\xi'\varrho,t|\xi(0)\varrho(0),\xi'(0)\varrho'(0),0), \\
\end{split}
\end{equation}
where
\begin{equation*} 
\begin{split}
    &\mathcal{J}(\xi\varrho,\xi'\varrho,t|\xi(0)\varrho(0),\xi'(0)\varrho'(0),0)) \\
    &\hspace{1cm}=\int\mathcal{D}\xi\mathcal{D}\xi'\mathcal{D}\varrho\mathcal{D}\varrho'\,e^{i(S_{\rm sys}[\xi,\varrho]-S_{\rm sys}[\xi',\varrho'])}e^{iS_{\rm IF}[\xi,\varrho,\xi',\varrho']},
\end{split}
\end{equation*}
with $e^{iS_{\rm IF}}$ given in Eq.~\eqref{Influence-functional-final} in terms of the variable $x_{ij}(t)$, defined in Eq.~\eqref{x-ij(t)-definition}, and
\begin{equation*}
    S_{\rm sys}[\xi,\varrho]=\frac{1}{2}\int \textrm{d} t\,m_0\delta_{ij}\dot{\xi}^i\dot{\xi}^j+\lambda\int \textrm{d} t\,\mathcal{L}(\varrho,\dot{\varrho}\Bar{t})-\frac{1}{2}\lambda\int \textrm{d} t\,\mathcal{L}(\varrho,\dot{\varrho}\Bar{t})\delta_{ij}\dot{\xi}^i\dot{\xi}^j.
\end{equation*}
The path integral over $\xi(t)$ is taken from $\xi(0)$ to $\xi$ and similarly for the other ones.

Following the discussion at the end of Sec. \ref{Sec:Gravitational-influence-functional}, within the approximation~\eqref{Influence-functional-final-approximated} we find
\begin{equation*}
\begin{split}
    &\mathcal{J}(\xi\varrho,\xi'\varrho,t|\xi(0)\varrho(0),\xi'(0)\varrho'(0),0)) \\
    &\hspace{1cm}=\int\mathcal{D}\xi\mathcal{D}\xi'\mathcal{D}\varrho\mathcal{D}\varrho'\mathcal{D}\mathcal{N}\,P[\mathcal{N}]\,e^{i(S_{\rm eff}[\xi,\varrho,\mathcal{N}]-S_{\rm eff}[\xi',\varrho',\mathcal{N}])},
\end{split}
\end{equation*}
where
\begin{equation*}
\begin{split}
    S_{\rm eff}[\xi,\varrho,\mathcal{N}]&=\int \textrm{d} t\,\left[ \frac{1}{2}m_0\delta_{ij}\dot{\xi}^i\dot{\xi}^j-\mathcal{N}_{ij}\xi^i\xi^j\right. \\
    &\hspace{1.5cm}\left.-\lambda\left( \frac{1}{2}\delta_{ij}\dot{\xi}^i\dot{\xi}^j-\frac{1}{m_0}\mathcal{N}_{ij}\xi^i\xi^j-1\right) \mathcal{L}(\varrho,\dot{\varrho}\bar{t}) \right] 
\end{split}
\end{equation*}
is an effective action describing the system whose dynamics are influenced by the stochastic variable $\mathcal{N}_{ij}(t)$.

Ultimately we are only interested in the reduced dynamics of the external DoFs of the composite particle. Therefore, we have to compute the reduced density matrix of the relevant degrees of freedom while tracing out all the others. Let us assume that initially the external and internal DoFs of our system are uncorrelated, thus implying that 
\begin{equation*}
    \rho_{\rm sys}\left(\xi(0)\varrho(0),\xi'(0)\varrho'(0),0\right) =\rho_{\rm ext}\left(\xi(0),\xi'(0),0)\rho_{\rm int}(\varrho(0),\varrho'(0),0\right),
\end{equation*}
where $\rho_{\rm ext}$ ($\rho_{\rm int}$) stands for the external (internal) DoFs density matrix. The reduced density matrix describing the external variables alone at time $t$ is obtained by taking the partial trace with respect to the internal DoFs
\begin{equation*}
    \rho_{\rm ext}(\xi,\xi',t)=\int \textrm{d}\varrho\,\rho_{\rm sys}(\xi\varrho,\xi'\varrho,t).
\end{equation*}
resulting in
\begin{equation*}
\begin{split}
    \rho_{\rm ext}(\xi,\xi',t)&=\int \textrm{d}\xi(0)\textrm{d}\xi'(0)\,\rho_{\rm ext}(\xi(0),\xi'(0),0) \\ &\hspace{0.5cm}\times\int\mathcal{D}\xi\mathcal{D}\xi'\mathcal{D}\mathcal{N}\,P[\mathcal{N}]\,e^{i(S_{\rm eff}^{(1)}[\xi,\mathcal{N}]-S_{\rm eff}^{(1)}[\xi',\mathcal{N}])}e^{S_{\rm IF}^{(\textrm{int})}[\xi,\xi',\mathcal{N}]},
\end{split}
\end{equation*}
where we have defined the new influence functional
\begin{equation} \label{Internal-DoF-IF}
\begin{split}
    e^{iS_{\rm IF}^{(\textrm{int})}[\xi,\xi',\mathcal{N}]}&=\int \textrm{d}\varrho\textrm{d}\varrho(0)\textrm{d}\varrho'(0)\,\rho_{\rm int}(\varrho(0),\varrho'(0),0) \\
    &\hspace{0.5cm}\times\int\mathcal{D}\varrho\mathcal{D}\varrho'\,e^{i(S_{\rm eff}^{(2)}[\xi,\varrho,\mathcal{N}]-S_{\rm eff}^{(2)}[\xi',\varrho',\mathcal{N}])},
\end{split}
\end{equation}
with
\begin{subequations}
\begin{equation}
    S_{\rm eff}^{(1)}[\xi,\mathcal{N}]=\int \textrm{d} t\left( \frac{1}{2}m_0\delta_{ij}\dot{\xi}^i\dot{\xi}^j-\mathcal{N}_{ij}\xi^i\xi^j\right) ,
\end{equation}
and
\begin{equation} \label{S-eff-(2)}
    S_{\rm eff}^{(2)}[\xi,\varrho,\mathcal{N}]=\lambda\int \textrm{d} t\left( 1-\frac{1}{2}\delta_{ij}\dot{\xi}^i\dot{\xi}^j+\frac{1}{m_0}\mathcal{N}_{ij}\xi^i\xi^j\right) \mathcal{L}(\varrho,\dot{\varrho}\bar{t}).
\end{equation}
\end{subequations}

Thus, we essentially have a similar problem to the one treated in Sec.~\ref{Sec:Gravitational-influence-functional}, namely a system interacting with a quantum environment, for which we shall compute the Feynman-Vernon influence functional once again. It is worth remarking that the total influence functional (gravitons plus internal DoFs) is not simply the product of the individual functionals, since the gravitational field couples with all variables describing the system. When considering the system of interest to be the external degrees of freedom, we effectively end up with two environments that interact with the system and with each other. In such a case, the additive property of the influence action for multiple environments does not hold.

In order to proceed, let us assume that the Lagrangian describing the internal degrees of freedom is of the form
\begin{equation*}
    \mathcal{L}(\varrho,\dot{\varrho}\,\bar{t})=\sum_\alpha\left[ \frac{1}{2}\mu_\alpha\Bar{\varrho}_\alpha^2-\mathcal{V}(\varrho_\alpha)\right] ,
\end{equation*}
with $\mu_{\alpha}$ representing reduced masses of the system and $\mathcal{V}$ being a function of the coordinates. Since we want to keep terms only up to second order in the position and velocity coordinates, we may write
\begin{equation*}
    \mathcal{L}(\varrho,\dot{\varrho}\,\bar{t})\simeq\sum_\alpha\left( \frac{1}{2}\mu_\alpha\dot{\varrho}_\alpha^2-\vartheta_\alpha\varrho_\alpha-\frac{1}{2}\mu_\alpha\varpi_\alpha^2\varrho_\alpha^2\right) ,
\end{equation*}
where $\vartheta_{\alpha}$ and $\varpi_{\alpha}$ are constants. Then, Eq.~\eqref{S-eff-(2)} becomes
\begin{equation*}
    S_{\rm eff}^{(2)}=\sum_\alpha\left[ \lambda\int \textrm{d} t\left( \frac{1}{2}\mu_\alpha\dot{\varrho}_\alpha^2-\frac{1}{2}\mu_\alpha\varpi_\alpha^2\varrho_\alpha^2\right) +\lambda\vartheta_\alpha\int\textrm{d} t\,Y(t)\varrho_\alpha(t)\right] ,
\end{equation*}
with
\begin{equation*}
    Y(t)=\frac{1}{2}\delta_{ij}\dot{\xi}^i\dot{\xi}^j-\frac{1}{m_0}\mathcal{N}_{ij}\xi^i\xi^j-1.
\end{equation*}

In this model, the internal DoFs, being represented by independent harmonic oscillators, are described by a quadratic action and also couple linearly with the external variables. In this case, the influence functional~\eqref{Internal-DoF-IF} is Gaussian, implying that we can write the influence action as
\begin{equation*}
\begin{split}
    S^{(\textrm{int})}_{\rm IF}[Y,Y']&=\int \textrm{d} t\textrm{d} t'\left\{ \frac{1}{2}[Y(t)-Y'(t)]D_{\rm int}(t,t')[Y(t')+Y'(t')]\right. \\
    &\hspace{1cm}\left. +\frac{i}{2}[Y(t)-Y'(t)]N_{\rm int}(t,t')[Y(t')-Y'(t')]\right\} ,
\end{split}
\end{equation*}
with
\begin{subequations}
    \begin{equation}
        D_{\rm int}(t,t')=i\lambda^2\sum_\alpha\vartheta_\alpha^2\langle\left[ \varrho_\alpha(t),\varrho_\alpha(t')\right] \rangle_{\rm int}\theta(t-t')
    \end{equation}
    and
    \begin{equation} \label{Internal-dofs-noise-kernel}
        N_{\rm int}(t,t')=\frac{1}{2}\lambda^2\sum_\alpha\vartheta_\alpha^2\langle\left\{ \varrho_\alpha(t),\varrho_\alpha(t')\right\} \rangle_{\rm int}
    \end{equation}
\end{subequations}
being the internal DoFs dissipation and noise kernels. Now the $\varrho(t)$'s are operators in the Heisenberg picture, and expectation values with the subscript 'int' are computed with respect to the initial state of the internal DoFs.

Similarly to what we did for the term containing the noise kernel for the gravitational influence functional, we can express the noise term in $S_{\rm IF}^{(\textrm{int})}$ in terms of a stochastic variable using the same Gaussian functional identity. Then, this term in the internal DoFs influence functional will give place to a Gaussian probability density and a linear term on the $Y$ variable. Therefore, just like in the gravitational case, the leading order contributions come from the noise term and we may write
\begin{equation}
    e^{iS_{\rm IF}^{(\textrm{int})}[Y,Y']}\simeq e^{-\frac{1}{2}\int dtdt'\,[Y(t)-Y'(t)]N_{\rm int}(t,t')[Y(t')-Y'(t')]},
\end{equation}
resulting in
\begin{align} \label{rho(xi,xi',t)}
    &\rho_{\rm ext}(\xi,\xi',t)=\int \textrm{d}\xi(0)\textrm{d}\xi'(0)\,\rho_{\rm ext}(\xi(0),\xi'(0),0) \int\mathcal{D}\xi\mathcal{D}\xi'\,e^{\frac{i}{2}m_0\delta^{ij}\int \textrm{d} t\,(\dot{\xi}_i\dot{\xi}_j-\dot{\xi}_i'\dot{\xi}_j')} \nonumber \\    &\hspace{0.5cm}\times\int\mathcal{D}\mathcal{N}\,P[\mathcal{N}]\,e^{-i\int \textrm{d} t\,\mathcal{N}^{ij}(\xi_i\xi_j-\xi_i'\xi_j')}e^{-\frac{1}{2}\int \textrm{d} t\textrm{d} t'\,[Y(t)-Y'(t)]N_{\rm int}(t,t')[Y(t')-Y'(t')]}.
\end{align}
The stochastic averages shown in Eqs.~\eqref{Stochastic-averages} can now be employed provided we work on a perturbative regime (dropping terms of order $\mathcal{O}(\xi^3)$ and higher). From this we obtain the external degrees of freedom density matrix as
\begin{align} \label{rho(xi,xi',t)-2}
    &\rho_{\rm ext}(\xi,\xi',t)=\int \textrm{d}\xi(0)\textrm{d}\xi'(0)\,\rho_{\rm ext}(\xi(0),\xi'(0),0)\int\mathcal{D}\xi\mathcal{D}\xi'\,e^{\frac{i}{2}m_0\delta^{ij}\int \textrm{d} t\,(\dot{\xi}_i\dot{\xi}_j-\dot{\xi}_i'\dot{\xi}_j')} \nonumber \\
    &\hspace{0.5cm}\times e^{-\frac{1}{4}\delta^{ij}\delta^{kl}\int \textrm{d} t\textrm{d} t'\,[\dot{\xi}_i(t)\dot{\xi}_j(t)-\dot{\xi}_i'(t)\dot{\xi}_j'(t)]N_{\rm int}(t,t')[\dot{\xi}_k(t')\dot{\xi}_l(t')-\dot{\xi}_k'(t')\dot{\xi}_l'(t')]} \nonumber \\
    &\hspace{0.5cm}\times e^{-\int \textrm{d} t\textrm{d} t'\,[\xi_i(t)\xi_j(t)-\xi'_i(t)\xi'_j(t)][\frac{1}{2}+m_0^{-2}N_{\rm int}(t,t')]N^{ijkl}_{\rm g}(t,t')[\xi_k(t')\xi_l(t')-\xi'_k(t')\xi'_l(t')]}.
\end{align}

As we can see, the off-diagonal elements of this density matrix in the coordinate basis ($\xi\neq\xi'$) evolve with an exponentially decaying amplitude. Therefore, for an initial superposition state of the external degrees of freedom, the interaction with the internal degrees of freedom induced by the gravitons as well as the interaction with the gravitons themselves inevitably leads to a loss of quantum coherence. This will be the subject of the next section. For now, let us compute an explicit form for the internal DoFs noise kernel.

\subsection{Internal DoFs noise kernel}

Since the free Lagrangian for each $\varrho_\alpha$ is the one of a harmonic oscillator, we may write the position operators in the Heisenberg picture as
\begin{equation*}
    \varrho_\alpha(t)=\sqrt{\frac{1}{2\mu_\alpha\varpi_\alpha}}(b_\alpha e^{-i\varpi_\alpha t}+b_\alpha^\dagger e^{i\varpi_\alpha t}),
\end{equation*}
where the $b$'s ($b^\dagger$'s) are annihilation (creation) operators satisfying the usual commutation relations. A direct calculation yields
\begin{equation*}
\begin{split}
    \langle\left\{ \varrho_\alpha(t),\varrho_\alpha(t')\right\} \rangle_{\rm int}&=\frac{2}{\mu_\alpha\varpi_\alpha^2}\langle H_\alpha\rangle_{\rm int}\cos\varpi_\alpha(t-t') \\
    &\hspace{0.2cm}+\frac{1}{\mu_\alpha\varpi_\alpha}\left[ \langle b_\alpha^2\rangle_{\rm int}e^{-i\varpi_\alpha(t+t')}+\langle(b^\dagger_\alpha)^2\rangle_{\rm int}e^{i\varpi_\alpha(t+t')}\right] ,
\end{split}
\end{equation*}
where $H_\alpha$ is the free Hamiltonian operator of the $\alpha$-th oscillator with frequency $\varpi_\alpha$.

Let us take the initial internal state to be a thermal one with temperature $\beta^{-1}$, so that $\langle b_\alpha^2\rangle_{\rm int}=\langle(b^\dagger_\alpha)^2\rangle_{\rm int}=0$. The canonical partition function reads
\begin{equation*}
    Z=\sum_{n=0}^\infty\exp\left[ -\varpi_\alpha\beta\left( n+\frac{1}{2}\right) \right] =\frac{e^{-\varpi_\alpha\beta/2}}{1-e^{-\varpi_\alpha\beta}},
\end{equation*}
and therefore
\begin{equation*}
    \langle H_\alpha\rangle_{\rm int}=\frac{\varpi_\alpha}{2}\coth\left( \frac{\varpi_\alpha\beta}{2}\right) .
\end{equation*}
We are then left with
\begin{equation*}
    \langle\left\{\varrho_\alpha(t),\varrho_\alpha(t')\right\} \rangle_{\rm int}=\frac{1}{\mu_\alpha\varpi_\alpha}\coth\left( \frac{\varpi_\alpha\beta}{2}\right) \cos\varpi_\alpha(t-t'),
\end{equation*}
and Eq.~\eqref{Internal-dofs-noise-kernel} becomes
\begin{equation*}
    N_{\rm int}(t,t')=\frac{\lambda^2}{2}\sum_\alpha\frac{\vartheta_\alpha^2}{\mu_\alpha\varpi_\alpha}\coth\left( \frac{\varpi_\alpha\beta}{2}\right) \cos\varpi_\alpha(t-t').
\end{equation*}

Let us further assume that the internal frequencies span a continuum, so we may replace
\begin{equation*}
    \sum_\alpha\to\int_0^\infty \textrm{d}\varpi\,\sigma(\varpi),
\end{equation*}
where $\sigma(\varpi)\textrm{d}\varpi$ is the number of oscillators with frequencies between $\varpi$ and $\varpi+\textrm{d}\varpi$. Then
\begin{equation*}
    N_{\rm int}(t,t')=\frac{\lambda^2}{2}\int_0^\infty\frac{\textrm{d}\varpi}{\varpi}\Omega(\varpi)\coth\left( \frac{\varpi\beta}{2}\right) \cos\varpi(t-t'),
\end{equation*}
with
\begin{equation*}
    \Omega(\varpi)\equiv\frac{\sigma(\varpi)\vartheta^2(\varpi)}{\mu_\varpi}.
\end{equation*}
If we consider the internal degrees of freedom to represent an Ohmic bath, we obtain~\cite{Calzetta2008}
\begin{equation*}
    \Omega(\varpi)=\gamma\varpi^2
\end{equation*}
for some coupling constant $\gamma$. For that case, we have
\begin{equation} \label{Int-noise}
    N_{\rm int}(t,t')=\frac{1}{2}\lambda^2\gamma\int_0^\infty \textrm{d}\varpi\,\varpi\coth\left( \frac{\varpi\beta}{2}\right) \cos\varpi(t-t').
\end{equation}
At high temperatures, $\beta<<t-t'$, the integral is dominated by low frequencies, and we find
\begin{equation} \label{Int-noise-high-T}
    N_{\rm int}(t,t')=\frac{\lambda^2\pi\gamma}{\beta}\delta(t-t'),
\end{equation}
namely a white noise.

\section{Decoherence induced by gravitons}

We now assume the system to be initially in a superposition state of two spatially separated locations
\begin{equation*}
    \rho_{\rm ext}(0)=|\Psi(0)\rangle\langle\Psi(0)|\hspace{0,5cm}\textrm{with}\hspace{0,5cm}|\Psi(0)\rangle=\frac{1}{\sqrt{2}}(|\boldsymbol{\xi}^{(1)}(0)\rangle+|\boldsymbol{\xi}^{(2)}(0)\rangle),
\end{equation*}
where $\boldsymbol{\xi}^{(1)}(t)$ and $\boldsymbol{\xi}^{(2)}(t)$ denote two classical distinguishable possible trajectories for the system. Let us consider that the superposition state survives for a time in the interval $0<t<t_f$ and that $\boldsymbol{\xi}^{(1)}(t)=\boldsymbol{\xi}^{(2)}(t)$ holds for $t\notin [0,t_f]$. We say that the coherence between the two components of this state will be effectively lost when $\exp[-\Gamma(t_f)]<<1$, where $\Gamma(t_f)$ is the decoherence rate~\cite{Kanno2021}. Such a rate can be estimated from Eq.~\eqref{rho(xi,xi',t)-2}, resulting in
\begin{equation} \label{Grav-decoherence-rate-1}
\begin{split}
    \Gamma(t_f)&=\delta^{ij}\delta^{kl}\int_0^{t_f}dtdt'\,V_i(t)\Delta v_j(t)N_{\rm int}(t,t')V_k(t')\Delta v_l(t') \\
    &+2\int_0^{t_f}dtdt'\,\Xi_i(t)\Delta\xi_j(t)\left[ 1+\frac{1}{m_0^2}2N_{\rm int}(t,t')\right] N^{ijkl}_{\rm g}(t,t')\Xi_k(t')\Delta\xi_l(t'),
\end{split}
\end{equation}
where
\begin{equation*}
    \Xi_i\equiv\frac{1}{2}(\xi_i^{(1)}+\xi_i^{(2)}),\hspace{0.5cm}\Delta\xi_i\equiv\xi_i^{(1)}-\xi_i^{(2)},
\end{equation*}
and similarly for $V_i$ and $\Delta v_i$. The quantities $V_i$ and $\Xi_i$, which are averages of the paths in superposition, are assumed to be time independent for simplicity~\cite{Kanno2021}.

We may now plug in the expression for the internal DoFs noise kernel into Eq.~\eqref{Grav-decoherence-rate-1}, which we found to be given by Eq.~\eqref{Int-noise-high-T} in the high temperature limit. For the decoherence rate, we are then left with
\begin{equation} \label{Grav-decoherence-rate-2}
\begin{split}
    \Gamma(t_f)&=\frac{\lambda^2\pi\gamma}{\beta}\delta^{ij}\delta^{kl}V_iV_k\int_0^{t_f}\textrm{d}t\,\Delta v_j(t)\Delta v_l(t) \\
    &+2\Xi_i\Xi_k\int_0^{t_f}\textrm{d}t\textrm{d}t'\,\Delta\xi_j(t)N_{\rm g}^{ijkl}(t,t')\Delta\xi_l(t') \\
    &+\frac{4\lambda^2\pi\gamma}{\beta m_0^2}\Xi_i\Xi_k\int_0^{t_f}\textrm{d}t\,\Delta\xi_j(t)N_{\rm g}^{ijkl}(t)\Delta\xi_l(t),
\end{split}
\end{equation}
where $N_{\rm g}^{ijkl}(t)\equiv\lim_{t'\to t}N_{\rm g}^{ijkl}(t,t')$.

Let us now consider a specific configuration of the superposition state. Take the separation $\Delta\xi_i(t)=\xi_i^{(2)}(t)-\xi_i^{(1)}(t)$ to be given by~\cite{Kanno2021,Breuer2001}
\begin{subequations} \label{Configuration-superposition-state}
\begin{equation}
    \Delta\xi_i(t)=\left\{
    \begin{array}{ll}
        2v_it &\textrm{for}\hspace{0.2cm}0<t\leq t_f/2\\
        2v_i(t_f-t) &\textrm{for}\hspace{0.2cm}t_f/2<t<t_f
    \end{array}
    \right.,
\end{equation}
such that
\begin{equation}
    \Delta v_i(t)=\left\{
    \begin{array}{ll}
        2v_i &\textrm{for}\hspace{0.2cm}0<t\leq t_f/2\\
        -2v_i &\textrm{for}\hspace{0.2cm}t_f/2<t<t_f
    \end{array}
    \right..
\end{equation}
\end{subequations}
For this configuration, the decoherence rate~\eqref{Grav-decoherence-rate-2} turns out to be
\begin{equation} \label{Grav-decoherence-rate-3}
\begin{split}
    \Gamma(t_f)&=8\Xi_iv_j\Xi_kv_l\left[ \int_0^{t_f/2}\textrm{d}t\textrm{d}t'\,tt'N_{\rm g}^{ijkl}(t,t')\right. \\
    &\hspace{1.0cm}+\int_{t_f/2}^{t_f}\textrm{d}t\textrm{d}t'\,(t_f-t)(t_f-t')N_{\rm g}^{ijkl}(t,t') \\
    &\hspace{1.0cm}\left. +2\int_0^{t_f/2}\textrm{d}t\int_{t_f/2}^{t_f}\textrm{d}t'\,t(t_f-t')N_{\rm g}^{ijkl}(t,t')\right] \\
    &\hspace{0.2cm}+\frac{16\lambda^2\gamma\pi}{\beta m_0^2}\Xi_iv_j\Xi_kv_l\left[ \int_0^{t_f/2}\textrm{d}t\,t^2N_{\rm g}^{ijkl}(t)+\int_{t_f/2}^{t_f}\textrm{d}t\,(t_f-t)^2N_{\rm g}^{ijkl}(t)\right].
\end{split}
\end{equation}

For gravitons initially in the vacuum state, the noise kernel is given by Eq.~\eqref{Noise-kernel-Minkowski-vacuum}. One can easily check that $\lim_{x\to0}F(x)=1/6$ and thus
\begin{equation*}
    N_{\textrm{g (vac)}}^{ijkl}(t)=\frac{1}{6}\frac{m_0^2\Lambda_{\rm g}^6}{15\pi}\mathcal{P}^{ijkl}.
\end{equation*}
The decoherence rate for the vacuum initial graviton state is then given by
\begin{equation} \label{Decoherence-rate-Minkowski-vacuum}
    \Gamma_{\rm (vac)}(t_f)=\frac{8m_0^2}{5\pi}\Lambda_{\rm g}^2\mathcal{K}\left[ G(\Lambda_{\rm g}t_f)+\lambda^2\kappa(\Lambda_{\rm g}t_f)^3\right] ,
\end{equation}
where
\begin{equation*}
    G(x)\equiv1+\frac{2}{3x}\left[ \sin x-8\sin\left( \frac{x}{2}\right) \right] +\frac{1}{x^2}\left[ \frac{2}{3}\cos x-\frac{32}{3}\cos\left( \frac{x}{2}\right) +10\right] ,
\end{equation*}
and $\mathcal{K}\equiv\mathcal{P}^{ijkl}\Xi_iv_j\Xi_kv_l$, while
\begin{equation*}
    \kappa\equiv\frac{\gamma\pi}{108m_0^2}\frac{\Lambda_{\rm g}}{\beta}.
\end{equation*}
We observe that when $\lambda=0$ (no internal degrees of freedom), we recover the same qualitative behavior found in Ref.~\cite{Kanno2021}, as expected.

\begin{figure}[!h]
    \centering
    \includegraphics[width=0.5\linewidth]{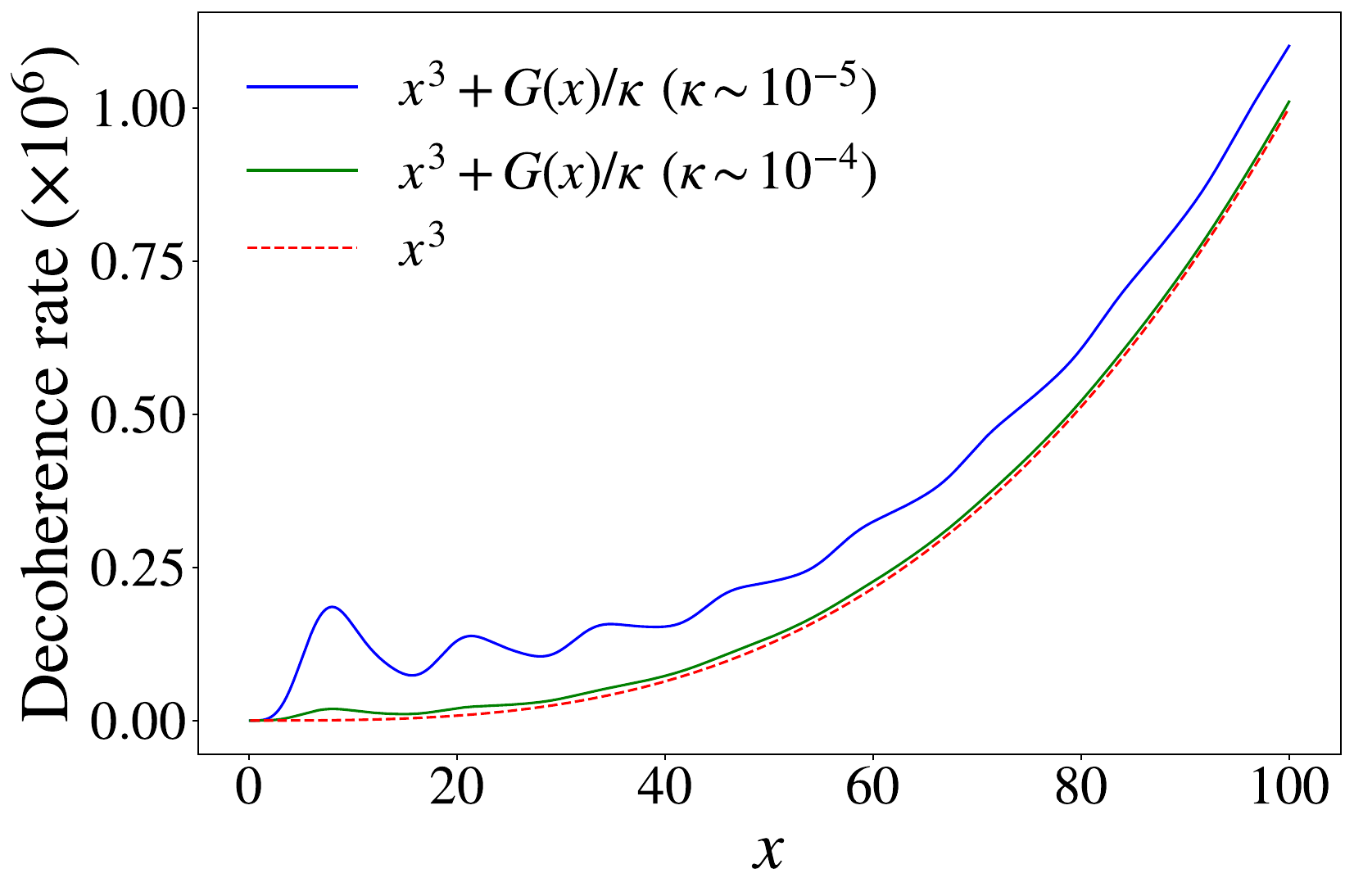}
    \caption{\textbf{Vacuum state}. Decoherence rate coming from the purely gravitational fluctuations and due to the presence of the internal degrees of freedom of the system. This last contribution becomes very important for $\kappa\gtrsim10^{-4}$.}
    \label{vac}
\end{figure}

The decoherence rate consists of the sum of two terms, one involving the decoherence induced exclusively by the weak quantum gravitational field and another one that comes from the coupling with both the internal DoFs and the gravitons. The intensity of the latter is measured by a constant $\kappa$. We expect this term to result in a greater contribution to the decoherence rate since it involves the coupling with two interacting environments, thus containing a further level of coarse-graining. In fact, Fig. \ref{vac} shows how, even for small values of $\kappa$, the behavior of the decoherence rate is dominated by the term involving not only the gravitons but also the internal degrees of freedom of the particle.

In regimes where the gravitons+internal DoFs contribution dominates, one may drop the exclusively gravitational contribution and, by setting $\lambda=1$ and restoring the constants $\hbar$, $c$, $k_B$ and $G$, the decoherence rate becomes
\begin{equation} \label{Decoherence-rate-Minkowski-vacuum-int+g}
    \Gamma_{\rm (vac)}(t_f)\simeq\frac{2\mathcal{K}}{135}\frac{Gk_B}{\hbar^4c^5}\gamma T\Lambda_{\rm g}^6t_f^3,
\end{equation}
where $T=(k_B\beta)^{-1}$.

The decoherence time is obtained by setting $\Gamma(\tau_{\rm dec})=1$. From Eq.~\eqref{Decoherence-rate-Minkowski-vacuum-int+g}, we find
\begin{equation}
    \tau_{\rm dec}^{\textrm{(vac)}}=\left( \frac{135}{2\mathcal{K}}\frac{\hbar^4c^5}{Gk_B}\frac{1}{\gamma T\Lambda_{\rm g}^6}\right) ^{1/3}.
\end{equation}
Let us note that the decoherence time does not depend on the mass of the composite particle, contrary to what we would have obtained by considering the exclusively gravitational contribution to the decoherence function. In fact, the decoherence time obtained in Ref.~\cite{Kanno2021} is proportional to the ratio $M_{\rm pl}/m_0$, where $M_{\rm pl}=\sqrt{\hbar c/G}$ is Planck mass. This means that, for a point particle, decoherence is not expected to occur for $m_0<<M_{\rm pl}\sim10^{-8}$ kg. However, when the system is also described by dynamical internal degrees of freedom, their coupling with gravity cannot be disregarded, since, as we have shown, it leads to a suppression of the purely graviton induced decoherence effect.

\section{Concluding remarks}

We considered the evolution of the external degrees of freedom of a freely falling particle under the action of two sources of noise: $i$) the coupling with its internal degrees of freedom and $ii$) the coupling with the quantized weak gravitational field. Due to the universal nature of the gravitational interaction, both environments are not independent, for which the immediate consequence is the non-additivity of the noise effects, resulting in a contribution to the decoherence rate that comes from the interaction between the two environments. This rate was found to be given by the sum of two terms: one arising purely from the quantum fluctuations of the gravitational field and another contribution that comes from the coupling with both the internal DoFs and the gravitons, whose intensity depends on the initial state of the gravitational field. We found that the presence of the dynamical internal DoFs causes a reduction in the decoherence time with respect to the one estimated in the literature by considering only the coupling with the gravitons. 

These results show that it is fundamental to take into account the quantum nature of matter as well as the quantum fluctuations of the gravitational field when considering phenomena like entanglement induced by gravity~\cite{Bose2017,Marletto2017,Danielson_2022,Christodoulou2023,Christodoulou2023b}, quantum reference frames~\cite{Rovelli1991,Giacomini2019} or decoherence induced by the gravitational time-dilation~\cite{Pikovski2015}, since some of these effects may be completely blurred by the loss of quantum coherence induced by the coupling of internal DoFs with gravity.

Furthermore, when computing the decoherence time by neglecting the much smaller purely gravitational contribution, we found that it does not depend on the mass of the system. This result could indicate that gravity might not be the ultimate explanation for the quantum-to-classical transition, which has been proposed in the literature~\cite{Bassi_2017} since the "collapse of the wave function" is expected to scale with the mass of the system. Although the purely gravitational contribution to decoherence does depend on the ratio $M_{\rm pl}/m_0$, the coupling with the internal DoFs, which suppresses the latter, does not.

\section*{Acknowledgments}
This work was supported by the National Institute for the Science and Technology of Quantum Information (INCT-IQ), Grant No.~465469/2014-0, by the National Council for Scientific and Technological Development (CNPq), Grants No.~308065/2022-0 and by the Coordination of Superior Level Staff Improvement (CAPES).

\renewcommand{\bibname}{References}
\begingroup
\let\cleardoublepage\relax

\endgroup

\end{document}